\documentclass[prd,aps,preprintnumbers,nofootinbib,superscriptaddress]{revtex4}
\pdfoutput=1
\usepackage{amssymb,amsmath,graphicx,bm}
\usepackage[final]{hyperref}
\usepackage{subfigure}

\usepackage[usenames,dvipsnames]{color}
\usepackage[normalem]{ulem} % \sout{old text} for strikeout
\renewcommand\sout{\bgroup \color[rgb]{0.55,0.00,0.99} \ULdepth=-.5ex \ULset}

\begin{document}
	
	\title{
Probing gluon GTMDs through exclusive coherent diffractive processes
 %\old{GTMD model predictions for exclusive coherent diffractive $J/\psi$ production}
 }
	
	\author{Dani\"el Boer}
	\email{d.boer@rug.nl}
	\affiliation{ {Van Swinderen Institute for Particle Physics and Gravity, University of Groningen, Nijenborgh 4, 9747 AG Groningen, The Netherlands}}
	
	\author{Chalis Setyadi}
	\email{c.setyadi@rug.nl; chalis@ugm.ac.id}
	\affiliation{ {Van Swinderen Institute for Particle Physics and Gravity, University of Groningen, Nijenborgh 4, 9747 AG Groningen, The Netherlands}}
	\affiliation{ {Department of Physics, Universitas Gadjah Mada, BLS 21 Yogyakarta, Indonesia
	}}
	
	\begin{abstract}
	We extend a previous GTMD model to improve the description of the HERA data on diffractive dijet production, and include exclusive coherent diffractive $J/\psi$ production data. We find that within our gluon GTMD model context and assumptions, there is considerable tension between the data for these two types of processes concerning the $t$ dependence. Photo- and electroproduction data for protons and nuclei from EIC and UPC data from LHC and RHIC can help to establish whether a common GTMD description is possible, as one would expect, and to facilitate studies of such data we provide predictions for the various experiments. We point out explicitly in which sense this goes beyond the description in terms of GPDs.	
    \end{abstract}
	
	%\pacs{12.38.-t; 13.85.Ni; 13.88.+e}
	\date{\today}
	
	\maketitle
	
\section{Introduction}
Coherent diffractive processes form a promising way to probe Generalized Transverse Momentum Dependent parton distributions (GTMDs). For example, exclusive coherent diffractive dijet production in electron-proton collisions has been suggested as a probe of gluon GTMDs in \cite{Hatta:2016dxp} (see also \cite{Mantysaari:2019csc}). By measuring the transverse momenta of the two jets one can access the off-forwardness as well as the transverse momentum distribution of gluons, albeit the latter only in an indirect way through a weighted integral where the weight is a function of the observed momenta (see e.g.\ \cite{Boer:2021upt}). Exclusive coherent diffractive $J/\psi$ production in electron-proton collisions can also be used \cite{Kowalski:2006hc,Bendova:2018bbb}, but in this case the weight of the integral cannot be varied substantially (double $J/\psi$ would be more suited for that), in that sense it is closer to processes like DVCS that probe GPDs, which are fixed integrals of GTMDs \cite{Hatta:2017cte}. Nevertheless, if the underlying GTMD description is valid, then the various processes should be describable simultaneously by the same GTMDs. There is data from the H1 and ZEUS experiments of HERA, which can be used to check this to a certain extent. In this paper we will attempt such a combined analysis and reach the  conclusion that there is considerable tension between the optimal dijet and $J/\psi$ descriptions. This tension can be due to the theoretical assumptions that go into the GTMD description, such as the selected GTMD model form or the model for the $J/\psi$ wave function, but can also have an experimental origin. This offers an excellent opportunity for the future U.S.-based Electron-Ion Collider (EIC) as it can provide additional and more precise data on both processes in different kinematical regions. Other tests of the underlying GTMD description can come from Ultra-Peripheral Collisions (UPCs) data of RHIC and/or LHC. Here it is important to ensure that the processes are exclusive and coherent diffractive which means the proton or nucleus has to remain intact. 
	
In \cite{Boer:2021upt} we have considered a McLerran-Venugopalan (MV) based (small $x$) model in order to obtain a description of diffractive dijet production data of H1. Although the data is neither fully exclusive, nor fully coherent, the kinematics is such that in leading order a GTMD description is expected to be appropriate. The model was indeed able to describe the $t$ dependence and the transverse momentum of the jets, but since it did not include any $x$ dependence, the $Q^2$ and $y$ dependence was not described very well. For the present study the first step will thus be to include an $x$ dependence, to obtain a better description of the $Q^2$ and $y$ dependence, before moving on to the description of $J/\psi$ production. We will show that the $x$-dependent GTMD model can provide a good description of the $t$ dependence of the exclusive coherent diffractive $J/\psi$ production data of H1 and ZEUS for different $Q^2$ values, including the photo-production case. 
However, it turns out there is no choice of parameters that leads a good description of the dijet and $J/\psi$ data {\it simultaneously}. The tension is solely in the slope of the exponential fall-off in $t$, which in the model is entirely governed by the width of the proton profile. We do no intend to resolve this tension in this paper, because there may be other reasons for the tension beyond the GTMD model and future data will be needed to confirm or refute the tension. 
We use the optimal model for the $J/\psi$ data to obtain predictions for the EIC and for UPCs, rather than the model that has the minimal tension with the dijet data as it would not yield a satisfactory description of either process.

In Sec.~\ref{sec:difdijet_GTMD} we collect the key ingredients of the diffractive dijet production cross section description in terms of the small-$x$ gluon GTMD of our previous study \cite{Boer:2021upt}. We then propose an $x$-dependent GTMD model to improve the description of the HERA-H1 data in Sec.~\ref{sec:x_dep_difdijet} and show the improved fit in Sec.~\ref{sec:difdijet_fit}. Our aim is to use as much as possible those fit parameters to also describe diffractive $J/\psi$ production data at HERA and LHC. The diffractive $J/\psi$ production cross section expression in terms of GTMDs is given in Sec.~\ref{sec:JPsi_intro}, followed by discussions about certain phenomenological corrections often considered. The fit results to H1 and ZEUS $\gamma^{(*)}p$ data are presented in Sec.~\ref{sec:JPsi_fit} where predictions to the future EIC are also provided. By further fitting ALICE Run 2 data on mid-rapidity UPCs off Pb nuclei (in order to include $A$ dependence in the model), we then give predictions for RHIC, LHC (at Run 1 center of mass energy), and EIC (for Au nuclei) in Sec.~\ref{sec:JPsi_pred}. We summarize our findings in Sec.~\ref{sec:discussion}.
	
\section{Diffractive Dijet Production in terms of GTMDs}
 \label{sec:difdijet_GTMD}
	
We first recall the essential expressions for the dijet case in terms of the small-$x$ gluon GTMD ${\cal F}_0^{[\Box]}(x,k_\perp, \Delta_\perp)$ \cite{Boer:2021upt}, which is the angular independent part of the more general GTMD
	\begin{equation}
		{\cal F}^{[\Box]}(x,\bm{k}_\perp, \bm{\Delta}_\perp) = {\cal F}_0^{[\Box]}(x,k_\perp, \Delta_\perp) + 2 {\cal F}_2^{[\Box]}(x,k_\perp, \Delta_\perp) \cos 2\theta_{k\Delta} + ...,
		\label{Fbox}
	\end{equation}
where $\theta_{k\Delta}$ denotes the angle between $\bm{k}_\perp$ and $\bm{\Delta}_\perp$. In this paper we will only consider the angular independent part ${\cal F}_0^{[\Box]}$, under the assumption that the contributions of the angular modulations (that enter the cross sections squared) are much smaller. For a detailed analysis of angular modulations (and shape fluctuations) we refer to \cite{Mantysaari:2020lhf}.
We also consider the off-forwardness to be entirely in the transverse direction, i.e.\ we consider zero skewness $\xi$.

In the strict $x\to 0$ limit and for zero skewness ($\xi=0$) one finds the following expression in terms of the dipole scattering amplitude ${\cal N}=1-\langle S^{[\Box]} \rangle_C$: 
	\begin{equation}
		{\cal F}^{[\Box]}(\bm{k}_\perp, \bm{\Delta}_\perp) = 4N_c \int \frac{d^2 \bm{r}_\perp d^2 \bm{b}_\perp}{(2\pi)^2} \, 
		e^{-i\bm{k}_\perp\cdot \bm{r}_\perp} e^{i\bm{\Delta}_\perp \cdot \bm{b}_\perp} \, \left[1-\langle S^{[\Box]}(\bm{b}_\perp,\bm{r}_\perp) \rangle_C \right].
		\label{Fbox2}
	\end{equation}
The average with subscript $C$ is over the color configurations of the proton or nucleus considered. The Wilson loop operator $S^{[\Box]}(\bm{x}_\perp,\bm{y}_\perp) \equiv {\rm Tr}\left[U^{[\Box]}(\bm{y}_\perp,\bm{x}_\perp)\right]/N_c$, where $U^{[\Box]}(\bm{y}_\perp,\bm{x}_\perp)=U^{[+]}(\bm{y}_\perp,\bm{x}_\perp) U^{[-]}(\bm{x}_\perp,\bm{y}_\perp)$ and $U^{[\pm]}$ are the standard staple-like gauge links in the forward ($+$) and backward ($-$) lightcone directions (see e.g.\ \cite{Bomhof:2006dp,Boer:2018vdi}). Furthermore, $\bm{r}_\perp=\bm{y}_\perp-\bm{x}_\perp$ and $\bm{b}_\perp=(\bm{x}_\perp+\bm{y}_\perp)/2$. The MV-like model used for $S^{[\Box]}$ will be discussed in the next section.
	
The $\gamma^* p$ differential cross section for $Q^2 > 0$ can be expressed in terms of amplitudes ${\cal A}_{\text{T}}$ and ${\cal A}_{\text{L}}$ for transverse and longitudinal photon polarizations, respectively:
	\begin{equation}
		\frac{d\sigma^{\gamma^* p \to j j p}_{T}}{d K_\perp  d\Delta_\perp^2} 
		=  \frac{(2 \pi)^4 \alpha_{em}}{16 N_c} \sum_f e_f^2 \ \int dz  \left[ z^2 +(1-z)^2 \right]  \, \frac{{\cal A}^2_{\text{T}} (K_\perp, \Delta_\perp, z, Q, y) }{K_\perp},
		\label{ATransv}
	\end{equation}
	and 
	\begin{equation}
		\frac{d\sigma^{\gamma^* p\to j j p}_L}{d K_\perp  d\Delta_\perp^2} 
		=   \frac{(2 \pi)^4 \alpha_{em}}{4 N_c} \sum_f e_f^2 \  \int dz \, z^2 (1-z)^2  
		\, \frac{{\cal A}^2_{\text{L}} (K_\perp, \Delta_\perp, z, Q, y) }{K_\perp},
		\label{ALong}
	\end{equation}
where $j$ denotes an outgoing jet, $y$ is the inelasticity, $e_f$ is the electric charge of a quark of flavor $f$ in units of the positron charge, $z=k_{1}^+/k^+$ is the outgoing quark (jet) longitudinal momentum fraction with respect to the virtual photon longitudinal momentum, $K_\perp = (k_{1\perp}-k_{2\perp})/2$ with $k_{i\perp}$ the  transverse momentum of jet $i$, $\Delta_\perp = k_{1\perp}+k_{2\perp}$, and $Q^2$ is the photon virtuality. The amplitudes ${\cal A}_{\text{T,L}}$ are given in terms of ${\cal F}_0^{[\Box]}$ as follows:
	\begin{equation}
		{\cal A}_{\text{T}} (K_\perp, \Delta_\perp, z, Q, y) = 
		\int \frac{d^2 \bm{q}_\perp}{(2 \pi)^3} 
		\left[\frac{ \bm{K}_\perp \cdot \left(\bm{K}_\perp -\bm{q}_\perp \right) }{z (1-z) Q^2 + \left(\bm{K}_\perp -\bm{q}_\perp \right)^2}\right] \left.\mathcal{F}^{[\Box]}_0 (x,q_\perp,{\Delta_\perp}) \right|_{x=s/(yQ^2)},
  \label{difdijet_T}
	\end{equation}
	and
	\begin{equation}
		{\cal A}_{\text{L}} (K_\perp, \Delta_\perp, z_i, Q)	= 
		\int \frac{d^2 \bm{q}_\perp}{(2 \pi)^3}  
		\left[\frac{ Q K_\perp}{z (1-z) Q^2 + \left(\bm{K}_\perp -\bm{q}_\perp \right)^2}\right]
		\left.\mathcal{F}^{[\Box]}_0 (x,q_\perp,{\Delta_\perp}) \right|_{x=s/(yQ^2)}.
  \label{difdijet_L}
	\end{equation}
	These expressions show that exclusive coherent diffractive dijet production allows to obtain information on GTMDs even though the transverse momentum dependence is integrated over. The dependence on the external momenta of the weights inside the integrals can be used to study the transverse momentum dependence of the GTMDs. For photoproduction $\gamma p$ one can set $Q^2 =0$ and drop the longitudinal part as it does not give any contribution.
 
	The expressions also show that exclusive coherent diffractive dijet production allows to obtain information on GTMDs that goes beyond the GPDs. In \cite{Hatta:2017cte} it was pointed out that the gluon GPD $H_g$ at small $x$ can be expressed in terms the small-$x$ gluon GTMD through the following integral\footnote{Taking into account that the definition of $\mathcal{F}^{[\Box]}_0$ differs from that of $F_0$ in \cite{Hatta:2017cte}, Eq.\ (\ref{GPDasintegral}) differs by a factor 2 from the one given in \cite{Hatta:2017cte}.}: 
	\begin{equation}
			xH_g (x, \Delta_\perp) = \frac{1}{(4 \pi)^2\alpha_s} \int d^2 \bm{q}_\perp \bm{q}_\perp^2 \mathcal{F}^{[\Box]}_0 (x,q_\perp,{\Delta_\perp}).
			\label{GPDasintegral}
	\end{equation}
	This relation is derived using the operator definitions of the functions, not taking into account renormalization.  
	However, this relation suffers from the same problems as its forward limit, where the collinear PDF $f(x)$ is viewed as the integral of a Transverse Momentum Dependent parton distribution (TMD): $f(x) \stackrel{?}{=} \int d^2 \bm{q}_\perp f(x,q_\perp^2)$ \cite{Collins:2011zzd}. Since the tail of the TMD $f(x,{q}_\perp^2)$ behaves as $1/q_\perp^2$, the integral will diverge logarithmically and requires some form of regularization. More formally, beyond tree level TMD factorization implies that the TMD depends on two scales, the rapidity scale $\zeta$ and the renormalization scale $\mu$, satisfying two coupled evolution equations, whereas the collinear PDF only depends on the scale $\mu$ and satisfies just one evolution equation. Clearly, Eq.\ (\ref{GPDasintegral}) suffers from the same problems. Because of this, here we do not provide curves for GPDs based on the GTMDs we obtain, as they would unavoidably depend on the adopted procedure of how to regulate the large transverse momentum behavior to make the integral in Eq.\ (\ref{GPDasintegral}) converge. Rather than expecting that the TMD determines the collinear PDF through an integral relation, one can instead consider the unambiguous relation in which the collinear PDF determines the large transverse momentum dependence of the TMD, i.e.\ $f(x,{q}_\perp^2) \propto \alpha_s q_\perp^{-2} \left(\int \frac{dy}{y} P(\frac{x}{y}) f(y) \right)$, where $P$ denotes a splitting function (ignoring for simplicity the possibility of mixing among various pdfs). Similar expressions hold for the perturbative large transverse momentum tails of GTMDs in terms of GPDs, as recently studied at the one loop level in \cite{Bertone:2022awq}. So rather than using fits of GTMDs to obtain results for GPDs, it is better to use models, lattice determinations, or fits of GPDs to predict the tails of the GTMDs and compare those to GTMD fits. This we do not attempt here, as we are primarily concerned with the small transverse momentum region in the present paper.
	
\section{An $x$-dependent GTMD Model}
\label{sec:x_dep_difdijet}
In our earlier study of diffractive dijet production \cite{Boer:2021upt} we used a small-$x$ model without any $x$ dependence, because the expressions in terms of the Wilson loop operator were obtained in the strict $x \to 0$ limit \cite{Dominguez:2010xd,Dominguez:2011wm,Boer:2015pni,Boer:2018vdi}. The model was able to describe the H1 data on the differential cross sections as function of $t$ and $K_\perp$ quite well, for which all data have the same small average $x$ value. But the model did not describe well the data as a function of $Q^2$ for which the average $x$ value is different for each data point or each bin. 

To improve on this we now incorporate an $x$ dependence in the model by replacing the constant free model parameter $\chi$ by the following function of $x$ and $Q^2$:
	\begin{equation}
		\chi (x) = \bar{\chi}  \left( \frac{x_0}{x} \right)^\lambda,	
	\end{equation}
with $x_0=3\times 10^{-4}$ and $\lambda=0.29$ based on the model by Golec-Biernat and W\"usthoff (GBW) for the saturation scale \cite{Golec-Biernat:1998zce,Golec-Biernat:1999qor}. This value of $\lambda$ also turns out to allow for a reasonably good description of the $Q^2$ dependence of the H1 diffractive dijet electroproduction data (using $x=s/(yQ^2)$) for diffractive dijet production, but as we will see later $J/\psi$ production prefers a somewhat smaller value $\lambda=0.22$.

Our motivation to include this type of $x$ dependence in our model is the observation of geometric scaling behaviour of DIS $ep$ collisions data in the low $x$ and low $Q^2$ region at HERA \cite{Golec-Biernat:1998zce,Golec-Biernat:1999qor}. This feature of the data was well-described by a saturation scale of the form $Q_s^2(x) \sim x^{-\lambda}$ with $\lambda \approx 0.29$. Later, it was shown that the total cross-section of $\gamma^* p$ exhibited geometric scaling over a much wider range of $Q^2$ values (from $0.045$ to $450 \, \text{GeV}^2$) in the $x<0.01$ region of HERA data \cite{Stasto:2000er} (see also \cite{Gelis:2006bs,Caola:2008xr}). Similar scaling was also observed in diffractive DIS with a specific parameterization \cite{Marquet:2006jb} and in inclusive $eA$ processes \cite{Freund:2002ux}.
	
We note that apart from this new $Q^2$ dependence introduced through the kinematic relation between $x$ and $Q^2$, we do not introduce any additional $x$ and/or $Q^2$ dependence from QCD corrections. We also consider a fixed coupling constant. The reason for not including evolution is that the scale evolution of GTMDs has not been studied in full yet. Even for TMD evolution the interplay between the $x$ and $Q^2$ evolution is not yet fully clear (the scale evolution of TMDs and Sudakov resummation for TMD processes at small $x$ has been investigated in e.g.\ \cite{Mueller:2012uf,Xiao:2017yya,Zhou:2018lfq,Zheng:2019zul}). Since the kinematic range of EIC is not too different from the one of H1 and ZEUS, we expect evolution not to be essential for obtaining predictions. Given the large uncertainties in the model parameters we expect logarithmic corrections to be of minor importance at this stage. This also avoids the question of what precisely sets the hard scale in the various processes (dijet versus $J/\psi$, electroproduction versus photoproduction). But the kinematic relation between $x$ and $Q^2$ for the $J/\psi$ case is affected by the $J/\psi$ mass, of course.

With all these considerations taken into account we arrive at the following McLerran-Venugopalan (MV)  \cite{McLerran:1993ni,McLerran:1993ka,McLerran:1994vd} like model for the dipole scattering amplitude:
	\begin{align}
		1-\langle S^{[\Box]}(\bm{b}_\perp,\bm{r}_\perp) \rangle_C
		=  \left[1- 
		\exp \left(-\frac{1}{4} {r}_\perp^2 \chi(x) Q_s^2(b_\perp) \ln \left[\frac{1}{{r}_\perp^2\Lambda^2} +e\right]\right)\right] e^{- \epsilon_r r^2_\perp}.
		\label{modelSbox}
	\end{align}
	Here the factor $e^{- \epsilon_r r^2_\perp}$ is introduced to ensure that the dipole sizes probed are restricted to the perturbative region. This Gaussian weight factor was also introduced in Wigner, Husimi and GTMD distributions \cite{Hagiwara:2016kam,Hagiwara:2017fye}.  In practice this should be enforced by the kinematics of the process, i.e.\ by the large transverse momentum of the jets or the large mass of the produced quarkonium, but by enforcing it explicitly in the model, we can obtain convergent integrals of the GTMD without reference to a process.
 In principle, $\epsilon_r$ is a free parameter of the model but we will use the fixed value $\epsilon_r=\left( 0.4 \ \text{fm} \right)^{-2}$ for both diffractive dijet and $J/\psi$ production, which corresponds to the gluonic radius of the proton used in \cite{Salazar:2019ncp}.
The free parameter $\bar{\chi}$ is chosen such as to obtain a reasonable description of the data.
	
	The above corresponds to a leading order description of a dipole interacting with a proton or nucleus through two-gluon exchange in the $t$-channel. The resulting $\langle S^{[\Box]}\rangle_C$, ${\cal N}$, and  ${\cal F}_0^{[\Box]}$ are purely real. In principle GTMDs can be complex, with an imaginary part that is referred to as the odderon contribution. In the forward limit for unpolarized protons this contribution has to vanish for the dipole case, but odderon contributions may arise for nuclei or from quadrupoles or higher multipoles. Even when the odderon contribution is considered absent down to a certain small $x$ value, nonlinear QCD evolution would 
 generate a nonzero contribution for even smaller $x$ values \cite{Hatta:2005as}. Therefore, in principle the imaginary part has to be included, but that has not been done in diffractive dijet production thus far. Later on we comment on the size of the expected correction from the imaginary part.
	
The saturation scale in the model will be taken as 
	\begin{align}
		Q_s^2(x,\bm{b}_\perp)  \equiv  \chi(x) Q_s^2(b_\perp) 
		= \bar{\chi} \left( \frac{x_0}{x}\right)^\lambda \frac{4 \pi \alpha_s^2  C_F }{ N_c} Q_{0s}^2 \, T(\bm{b}_\perp)
		\label{Qs2x}
	\end{align}
	with $T(\bm{b}_\perp)$ is the profile of the proton or nucleus. For the proton we use a Gaussian profile
	\begin{align}
		T_p (\bm{b}_\perp) = \exp \left( -\frac{\bm{b}_\perp^2}{2 R_p^2} \right)
  \label{proton_profile}
	\end{align}
	with $R_p$ the gluonic radius of the proton \cite{Salazar:2019ncp}. For heavy nuclei the profile is described by the thickness function \cite{Iancu:2017fzn}
	\begin{equation}
		T_A (\bm{b}_\perp ) = N_A \int dz \ \rho_A \left(  \sqrt{\bm{b}_\perp^2 + z^2} \right)
	\end{equation}
	which is obtained from the Woods-Saxon distribution
	\begin{equation}
		\rho_A(r) = \frac{1}{1 + \exp \left[ \frac{r-R_A}{a_A} \right]}.
	\end{equation}
	Here, $N_A$ is a normalization factor such that $T_A (\bm{b}_\perp=0) = 1$. For the numerical calculations we choose the nuclear radius $R_A=1.12A^{1/3}$ fm with $A$ the mass number of the nucleus\footnote{Using a different nuclear radius will affect the $t$ distribution of the cross section, as shown in \cite{Mantysaari:2022sux}, where a larger radius corresponds to a steeper slope.}.
 For lead ($A=208$) we use $a_A=0.546$ fm and for gold ($A=197$) we use $a_A=0.535$ fm.
	For the proton we use $Q_{0s,p}^2 =  1 \ \text{GeV}^2$, while (similar to what was done in \cite{Salazar:2019ncp}) for the heavy nuclei $Q^2_{0s,\text{A}}$ can be obtained from the relation 
	\begin{align}
		\int d^2 \bm{b}_\perp \ Q^2_{0s,\text{A}} (x,\bm{b}_\perp)  = A^{\eta} \int d^2 \bm{b}_\perp \ Q_{0s,p}^2 (x,\bm{b}_\perp)
  \label{Saturation}
	\end{align}
to be evaluated at the same value of $x$. Here $\eta$ will be considered a free parameter to be fitted to the data. Explicitly, for heavy nuclei we have 
	\begin{equation}
		Q^2_{0s,\text{A}} = A^{\eta} Q_{0s,p}^2 \frac{\int d^2 \bm{b}_\perp T_p(\bm{b}_\perp)}{\int d^2 \bm{b}_\perp T_A(\bm{b}_\perp)}
	\end{equation}
Requiring $\int d^2 \bm{b}_\perp T_A(\bm{b}_\perp) \propto R_A^2 \propto A^{2/3} $ we obtain the following expression for the saturation scale
	\begin{equation}
		Q^2_{0s,\text{A}} \propto A^{\eta - \frac{2}{3}} Q_{0s,p}^2
		\label{Saturation2}
	\end{equation}
which will determine the saturation scale of the heavy nuclei. The general expectation is that $\eta$ should have a value near 1.
	
\section{Diffractive dijet electroproduction}
\label{sec:difdijet_fit}
	The diffractive dijet electroproduction cross section $ep\rightarrow e'jjp$ can be related to the $\gamma^*_{\text{T,L}}p\rightarrow jj p$ cross sections in Eqs.\ (\ref{ATransv})
	and (\ref{ALong}) as follows:
	\begin{equation}
		\frac{d\sigma^{ep\rightarrow e'jjp}}{dx dQ^2} = \frac{\alpha_{\text{em}}}{\pi x Q^2} \left[ \left( 1-y+\frac{y^2}{2} \right) \sigma^{\gamma^* p \rightarrow jjp}_{\text{T}} +
		\left( 1-y \right) \sigma^{\gamma^* p \rightarrow jjp}_\text{L}\right].
	\end{equation}
	In the calculation, we use $\Lambda=0.24$ GeV, $C_F=4/3$, $N_c=3$, $\alpha_{\text{em}}=1/137$, $N_f=4$, and fixed $\alpha_s = 0.3$.
	
With the above model and cross section expressions we fit the H1 data of \cite{H1:2011kou}. Although the data is for the production of at least two jets and not fully exclusive, a leading order description in the measured kinematic regime will be dominated by the production of two jets in any case. Also the data is not fully for coherent diffraction, but the $t$ values are so small and the kinematic cuts on the final state proton are such that the additional light hadrons that may be produced in the process are not expected to alter the process much compared to the coherent case. The data is consistent within errors with actual exclusive coherent diffractive dijet production data from ZEUS \cite{Abramowicz:2015vnu} (to be specific, this we checked for $d\sigma/d\beta$ at $\beta=x_B/x_{I\!\!P}=0.1$). The ZEUS data is less differential however and therefore not used here. Finally, the H1 data do not have zero skewness, in fact, it may typically be larger than the $x$ values (the average value for $x_{I\!\!P} \approx \xi$ is around $0.03$-$0.04$ \cite{H1:2011kou}, whereas the geometric average of the upper and lower value of the $x$ range is $10^{-3}$), meaning that the ERBL region is probed rather than the DGLAP one, cf.\ e.g.\ \cite{Diehl:2003ny}. These caveats concerning this data should be kept in mind, when we discuss the tension with the best GTMD model description of the $J/\psi$ production case (for which $\xi$ is also not exactly zero, of course). This is also one of the reasons why we do not attempt to find the best model fit that can describe both processes simultaneously. The purpose here is to demonstrate those features that the model fits have in common and those that are in tension, such that future experimental investigations can focus particular attention on these aspects.
	
\begin{figure}[htb]
\centering
\subfigure{\includegraphics[width=0.45\textwidth]{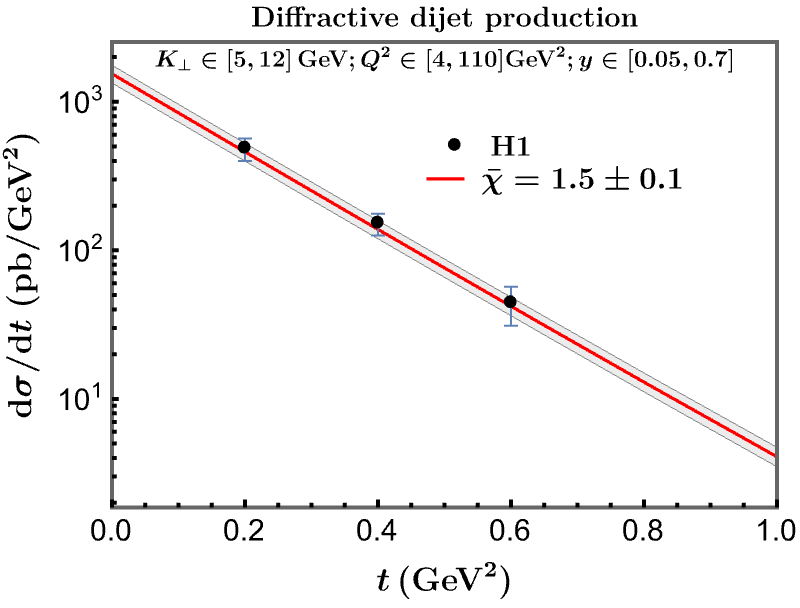}} 
\subfigure{\includegraphics[width=0.45\textwidth]{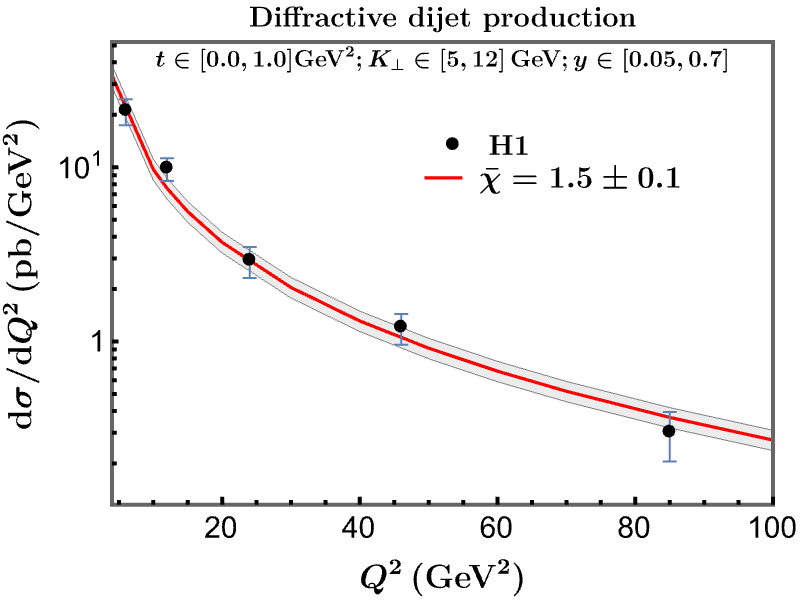}} 
\subfigure{\includegraphics[width=0.45\textwidth]{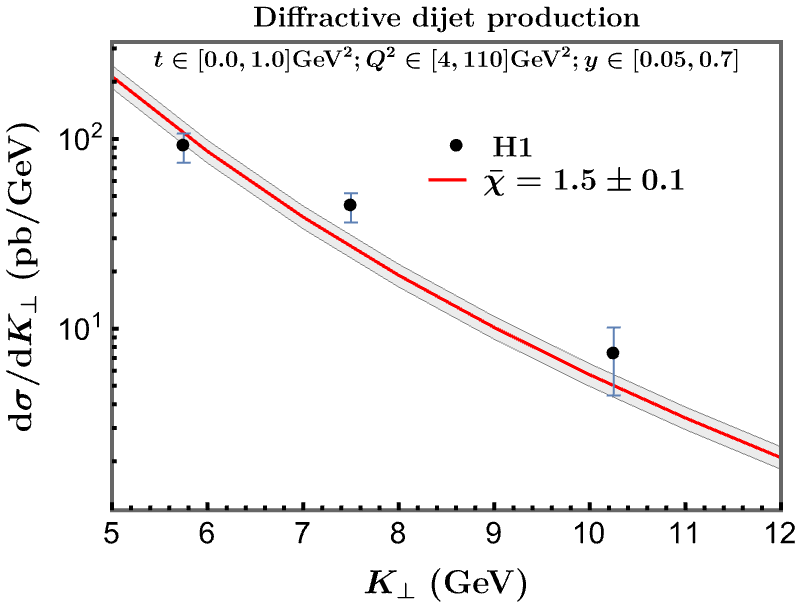}} 
\subfigure{\includegraphics[width=0.45\textwidth]{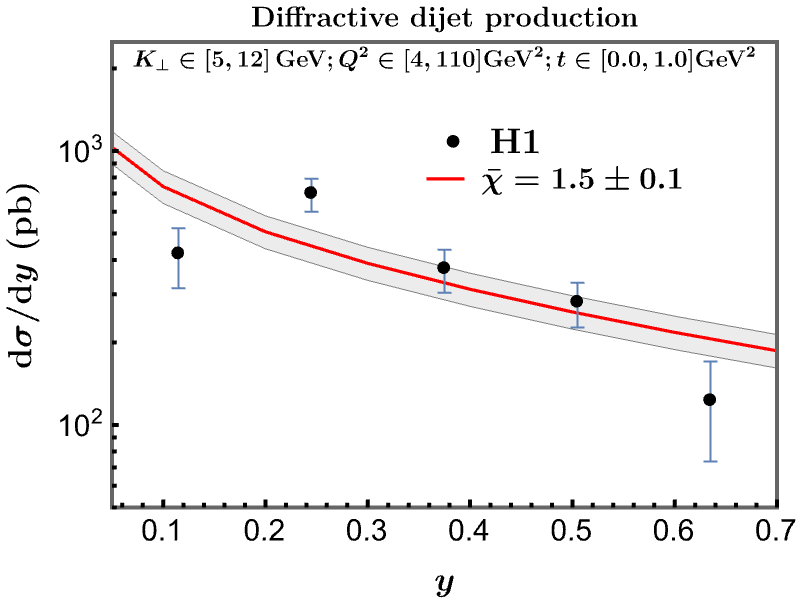}} 
\caption{The best fit of diffractive dijet production to H1 data with $R_p=0.49$ fm, $\lambda=0.29$, and $\epsilon_r=\left( 0.4 \ \text{fm} \right)^{-2}$. The central value corresponds to $\bar{\chi}=1.5$ while the band correspond to $\bar{\chi}=1.4$ and $\bar{\chi}=1.6$ for the lower and upper band respectively. The systematic and statistical uncertainties are added in quadrature.}
\label{DifDijet}
\end{figure}

The best fit to the $t$ distribution of H1 data \cite{H1:2011kou} is shown in Fig.\ \ref{DifDijet}.
The $t$ distribution is well described by an exponential fall-off, $d\sigma/dt \propto \exp (-bt)$, as also is the case for the $t$ distribution for $J/\psi$ production. The slope $b$ is mainly determined by the gluonic radius of the proton $R_p$ and we find that the model gives the best fit for $R_p=0.49\pm 0.02$ fm which leads to the slope $b=6.0 \pm 0.5 \, \text{GeV}^{-2}$ (in \cite{H1:2011kou} $b=5.89\pm 0.50 \, \text{GeV}^{-2}$ is given for the H1 data). 
Incorporating the statistical and systematic uncertainties of the data which are simply added in quadrature, we give a band around the best fit (central value) $\bar{\chi}=1.5\pm 0.1$.
Within errors, our model gives a reasonable description of the $Q^2$, $K_\perp$ and $y$ distribution data, an improvement w.r.t.\ the $x$ independent model of our previous study \cite{Boer:2021upt}.
	
\section{Exclusive coherent diffractive $J/\psi$ production}
\label{sec:JPsi_intro}

The differential cross section of exclusive coherent diffractive vector meson production as a function of momentum transfer squared $t=-\Delta_\perp^2$ (for $\xi=0$) can be written as \cite{Kowalski:2006hc,Watt:2007nr}
	\begin{equation}
		\frac{d \sigma_{T,L}^{\gamma^*  p \rightarrow Vp}}{dt} = \frac{1}{16 \pi} \left|  \mathcal{A}_{T,L}^{\gamma^*  p \rightarrow Vp}  \right|^2 \left(1+\beta_{T,L}^2 \right)R_g^2,
	\end{equation}
	expressed in terms of the elastic scattering amplitudes 
	\begin{equation}
		\mathcal{A}_{T,L}^{\gamma^*  p \rightarrow Vp} =
		2i \int d^2 {\bm{r}_\perp} \int d^2 {\bm{b}_\perp} \int_0^1 \frac{dz}{4\pi} \left(
		\Psi_V^* \Psi_{\gamma}
		\right)_{T,L} 
		\mathrm{e}^{-i \left[ {\bm{b}_\perp} -(\frac{1}{2}-z){\bm{r}_\perp} \right] \cdot {\bm{\Delta}_\perp}} \left[1-S(\bm{r}_\perp, \bm{b}_\perp)\right],
		\label{Scatt_A}
	\end{equation}
for the transverse/longitudinal polarization ($T/L$) of the virtual photon.
Here $\bm{b}_\perp$ denotes the impact parameter, $S(\bm{r}_\perp, \bm{b}_\perp)$ is (the real part of) the dipole-proton scattering amplitude, and $\left( \Psi_V^* \Psi \right)_{T,L}$ is the overlap between the photon and the vector meson wave functions which depends on the transverse dipole size $\bm{r}_\perp$, momentum fraction $z$ carried by the quark, the quark mass $m_f$, the vector meson mass $M_V$, and the photon virtuality $Q^2$. We are working in a frame where the colliding virtual photon and proton have zero transverse momentum. We note that in Eq.\ (\ref{Scatt_A}) we use the phase factor $e^{i(\frac{1}{2}-z)){\bm{r}_\perp} \cdot {\bm{\Delta}_\perp}}$ as suggested in \cite{Hatta:2017cte}, rather than $e^{i(1-z)){\bm{r}_\perp} \cdot {\bm{\Delta}_\perp}}$ as used in \cite{Kowalski:2006hc}. In practice, it does not make much difference though. In models studied by \cite{Bendova:2020hbb}, the cross section amplitude using the phase factor from \cite{Hatta:2017cte} is only 3.5\% larger than using the one from \cite{Kowalski:2006hc}.
	
As in the dijet case, the model parameter $R_p$ of the proton profile directly determines the $t$ slope of the differential cross section for $J/\psi$ production. However, the latter is found to be quite different from that of the dijet case, with the $J/\psi$ production slope substantially smaller. They are only compatible at the 3$\sigma$ level, hence there is considerable tension. Thus far no phenomenological study for both types of processes simultaneously has been performed and we will not attempt that here, but in principle both should be describable within the same GTMD based approach. Due to this aspect of tension, the uncertainty in the model parameters is larger than what is suggested by studying only one of the two types of process. Future data is needed to resolve this issue.
	
The photon-vector meson wave functions overlaps are expressed as \cite{Kowalski:2006hc}
	\begin{eqnarray}
		\left( \Psi^*_V \Psi
		\right)_T &=& \hat{e}_f e \frac{N_c}{\pi z (1-z)} \left[ m_f^2 K_0 (\epsilon \bm{r}_\perp) \phi_T (\bm{r}_\perp,z) - \left(z^2 + (1-z)^2 \right) \epsilon K_1 (\epsilon  \bm{r}_\perp) \partial_{\bm{r}_\perp} \phi_T (\bm{r}_\perp,z)  \right] \nonumber \\
		\left( \Psi^*_V \Psi
		\right)_L &=& 
		\hat{e}_f e \frac{N_c}{\pi } 2z (1-z) Q K_0 (\epsilon  \bm{r}_\perp) \left[ M_V  \phi_L (\bm{r}_\perp,z) + \delta \frac{m_f^2 -\nabla_{\bm{r}_\perp}^2}{M_V z(1-z)} \phi_L (\bm{r}_\perp,z)  \right] 
		\label{Overlap}
	\end{eqnarray}
where $\phi_{T,L}$ are model dependent scalar functions and $\delta$ is usually taken to be 0 or 1. Because the wave function overlap of the longitudinal part is linearly dependent on $Q$, the chosen value of $\delta$ does not affect photoproduction ($Q^2 \approx 0$). However, the chosen $\delta$ value significantly affects the large $Q^2$ cross section as the longitudinal part becomes a larger portion as $Q^2$ grows: the second term of $\left( \Psi^*_V \Psi \right)_L $ grows faster than the first term when $\delta=1$. In this study, we will use the so-called Gaus-LC (GLC) and boosted Gaussian (BG) vector meson models \cite{Kowalski:2006hc,Kowalski:2003hm}. 
The GLC model is given by:
\begin{eqnarray}
\phi_{T,L} (\bm{r}_\perp,z) &=& N_{T,L} z^2(1-z)^2  \exp \left[ -\frac{\bm{r}_\perp^2 }{2 R_{T,L}^2}  \right]. 
\end{eqnarray}
For $V=J/\psi$ the parameters are listed in \cite{Kowalski:2006hc}: $N_T=1.23$, $N_L = 0.83$, $R_T^2=6.5$ $\mathrm{GeV}^{-2}$, $R_L^2=3.0$ $\mathrm{GeV}^{-2}$, with $m_f = 1.4$ $\mathrm{GeV}$. The BG model is given by \cite{Nemchik:1994fp,Nemchik:1996cw}:
\begin{eqnarray}
\phi_{T,L} (\bm{r}_\perp,z) 
&=&
\mathcal{N}_{T,L} z(1-z) \exp \left[ a_1 (z) \bm{r}_\perp^2+ a_2 (z)\right]
\end{eqnarray}
with $a_1 (z)= - \frac{2z(1-z)}{\mathcal{R}^2} $ and $a_2 (z) =  -\frac{m_f^2 \mathcal{R}^2}{8z(1-z)} + \frac{m_f^2 \mathcal{R}^2}{2}$. 
The parameters are listed in \cite{Kowalski:2006hc} (see also \cite{Mantysaari:2018nng}):  
$\mathcal{N}_T = 0.578$, $\mathcal{N}_L = 0.575$, $\mathcal{R}^2=2.3 \, \mathrm{GeV}^{-2}$, and $M_V=3.097$ GeV.
For both models the parameters are fixed by requiring that the wave function is normalized and that the decay widths $f_V=f_{V,T}= f_{V,L}$ are reproduced. We note that other vector meson wave functions and potential models have been considered in order to describe diffractive production of $J/\psi$ and other heavy quarkonia, see for example \cite{Kopeliovich:1991pu,Eichten:1979ms,Eichten:1978tg,Quigg:1977dd,Cepila:2019skb,Barik:1980ai,Buchmuller:1980su,Kowalski:2003hm}. These descriptions can in principle also be translated into GTMD model expressions, which possibly allows for a more direct comparison of models (rather than a comparison of how they describe the data), but that is not our objective here.

Before moving on to the description of the data, we 
make some observations about whether this process probes a GPD rather than a GTMD. If we restrict to the angular independent part only, then we can express Eq.\ (\ref{Scatt_A}) as 
\begin{align}
\mathcal{A}_{T,L}
&= \frac{\pi i}{2 N_c} \int_0^1 dz \int d^2 {\bm{r}_\perp}  
\left(
\Psi_V^* \Psi_{\gamma}
\right)_{T,L} 
\left( \bm{r}_\perp,z	\right)
\int d^2 {\bm{q}_\perp} J_0 \left( | q_\perp + \delta_\perp| \bm{r}_\perp \right)
\mathcal{F}^{[\Box]}_0 (x,q_\perp,{\Delta_\perp}), 
\end{align}
with $\bm{\delta}_\perp = (\frac{1}{2} - z) \bm{\Delta}_\perp$. This expression shows that also in diffractive $J/\psi$ production one probes an integral over a GTMD with an integrand that depends on the kinematic variables of the process (in this case $z$ and $\Delta_\perp$) and hence different integrals of the GTMD can be obtained in this way, even though with less possibilities for varying the integrand than in the dijet case. This also means that the expression cannot be given in terms of a GPD (which does not depend on $z$), only in an approximation, as pointed out by \cite{Hatta:2017cte}. As the weight of the integral over the GTMD depends on $\Delta_\perp$ through $\delta_\perp$ and $\Delta_\perp$ is generally small and only relevant in a small kinematic region (and the region around $z=1/2$ contributes the most), this dependence may be ignored to good approximation leaving a fixed integral over the GTMD, which however still is not an expression in terms of the gluon GPD $H_g$ through Eq.\ (\ref{GPDasintegral}). This would require small $| q_\perp + \delta_\perp|$, for which we can consider the first order expansion of the Bessel function $J_0 \left( | q_\perp + \delta_\perp| \bm{r}_\perp \right) \approx 1 -\frac{(q_\perp + \delta_\perp)^2 r_\perp^2}{4}$, and use the fact that $\int d^2 \bm{q}_\perp \mathcal{F}^{[\Box]}_0 (x,q_\perp,{\Delta_\perp}) = 0$ \cite{Hatta:2017cte}, to find that
\begin{align}
\mathcal{A}_{T,L} & \approx \frac{\pi i }{8 N_c} \int_0^1 dz
\int d^2 {\bm{r}_\perp}  \bm{r}_\perp^2  
\left(
\Psi_V^* \Psi_{\gamma}
\right)_{T,L} 
\left( \bm{r}_\perp,z	\right)
\int d^2 {\bm{q}_\perp} \bm{q}_\perp^2 \mathcal{F}^{[\Box]}_0 (x,q_\perp,{\Delta_\perp}) \nonumber\\
& = \frac{\pi^3 i \alpha_s }{N_c} \int_0^1 dz
\int d^2 {\bm{r}_\perp} \bm{r}_\perp^2  
\left(
\Psi_V^* \Psi_{\gamma}
\right)_{T,L} 
\left( \bm{r}_\perp,z	\right) xH_g (x, \Delta_\perp), 
\end{align}
where in the last step we used Eq.\ (\ref{GPDasintegral}). It should be stressed that this is an approximation that depends on the relevant range of the $q_\perp$ integration. Therefore, we will consider the GTMD expression rather than the approximate one in terms of the GPD. Diffractive scattering in terms of GPDs has been studied in \cite{Braun:2005rg}. 
						
\section{Analysis of coherent diffractive $J/\psi$ production data}
\label{sec:JPsi_fit}
						
The total $\gamma^* p$ cross section for $J/\psi$ production studied by H1 \cite{H1:2005dtp} is defined as $\sigma_\textrm{tot} = \sigma_T + \varepsilon \sigma_L$ with $\varepsilon = (1-y)/(1-y+\frac{1}{2}y^2)$ (with $\langle \varepsilon \rangle  = 0.99$ in \cite{Kowalski:2006hc,Altinoluk:2015dpi}), while by ZEUS \cite{ZEUS:2004yeh} it is defined as $\sigma_\textrm{tot} = \sigma_T + \sigma_L$. Here we will also use $\varepsilon=0.99$, which corresponds to $\langle y \rangle \simeq 0.13$. This differs from the dijet case where a range of $y$ values is considered and the $x$ value depends on $y$. The $x$ value used in the $J/\psi$ production case is determined by \cite{Martin:1999wb,Watt:2007nr}
\begin{equation}
x =x_B\left( 1+\frac{M_V^2}{Q^2} \right)= \frac{M_V^2 + Q^2}{W^2+Q^2}.
\end{equation}
We recall that $\sigma_L=0$ for the photoproduction case.

Diffractive vector meson production at HERA and LHC have been studied extensively with various model approaches, such as \cite{Lappi:2010dd,Mantysaari:2018nng,Bendova:2020hbb,Cepila:2017nef,Cepila:2018zky,Guzey:2013qza,Goncalves:2014wna,Zhang:2021vgj,Watt:2007nr,Eskola:2022vpi}. It has been shown in \cite{Kowalski:2006hc,Armesto:2014sma,Bendova:2018bbb,Mantysaari:2020axf} that MV-like models can describe diffractive vector meson production well. In most model studies the dipole scattering amplitude is real and hence $\mathcal{A}_{T,L}^{\gamma^*  p \rightarrow Vp}$ imaginary, but as mentioned, in reality there will be an imaginary part. The phenomenological correction $1+\beta^2$ is introduced to account for that contribution. Using dispersion relations an expression for $\beta$ has been obtained \cite{Gribov:1968ie,Shuvaev:1999ce,Martin:1999wb}: 
\begin{equation}
\beta_{T,L} = \tan \left[ \frac{\pi \lambda_{T,L}}{2} \right], \qquad
\lambda_{T,L} \equiv \frac{\partial \ln \left[ \mathcal{A}_{T,L}^{\gamma^*  p \rightarrow Vp}  \right]}{\partial \ln \left[ \frac{1}{x} \right]}. 
\end{equation}
This expression implies that only $x$ dependent elastic scattering amplitudes yield nonzero $\beta$. Phenomenological studies of HERA data \cite{Martin:1999wb,Toll:2012mb,Mantysaari:2016jaz} find that the real part correction $(1+\beta^2)$ is anywhere between 10\% and 25\%. Given the large uncertainty in the model fits that we obtain, this correction will not be of much importance. Therefore, we will first fit the data without this correction and then estimate the size of the correction for that fit afterwards. In this way we find that 
in our case ($1+\beta^2$) is in the 10-15\% range. More details on this will be presented below.
						
Another correction often considered comes from taking into account nonzero skewness. The off-forwardness
$\Delta= P'-P$ 
for zero skewness, i.e.\ $\xi = - \Delta^+/({P'}^+ + P^+)=0$, means $\Delta=\Delta_\perp$ and $t=-|\bm{\Delta}_\perp^2|$. In practice $\xi$ will not be exactly zero. Therefore, in order to correct for this the factor $R_g^2$ is included, where $R_g$ for gluons at small $x$ and small $\xi$ is given by \cite{Shuvaev:1999ce}  
\begin{eqnarray}
R_g  &\equiv & \frac{H_g(x=\xi,\xi)}{H_g(x=2\xi,0)} \approx \frac{2^{2 \delta + 3}}{\sqrt{\pi}} \frac{\Gamma(\lambda + \frac{5}{2})}{\Gamma(\lambda +4)} ,
\end{eqnarray}
where $H_g(x,\xi)$ is the standard (helicity non-flip) gluon GPD (now for nonzero skewness, unlike in Eq.\ (\ref{GPDasintegral})). This factor $R_g^2$ is a substantial correction for HERA kinematics, found to be in the order of 40-70\% \cite{Martin:1999wb,Mantysaari:2016jaz,Toll:2012mb}. We note though that the value $H_g(x=\xi,\xi)$ is at the boundary of the ERBL and DGLAP regions, where the function is continuous but not differentiable, hence changing abruptly, so a slight change in $x$ w.r.t.\ $\xi$ or vice versa can matter considerably. This introduces an uncertainty regarding the actual correction that is needed, as the data span a range of $x$ and $\xi$ values. Furthermore, the specific value $x= 2\xi$, which is in the DGLAP region, stems from the GPD analysis of \cite{Flett:2019pux} for which the region around this value is found to make the dominant contribution. This is quite different from our GTMD approach for which $x= 2\xi$ plays no dominant role and the ratio $H_g(x,\xi)/H_g(x,0)$ (i.e.\ with the same $x$ value in numerator and denominator) for small nonzero $\xi$ seems more appropriate to consider. Moreover, as mentioned, the data are more likely in the ERBL region ($x < \xi$), although that is not clearly specified in the experimental papers. Hence, given these considerations, here we do not include the $R_g^2$ correction factor in this paper and also not correct for nonzero skewness in another way. Given the large uncertainties in the model fits (due to the large uncertainties in the data), this is not expected to be essential.

\begin{figure}[htb]
\centering
\subfigure{\includegraphics[width=0.475\textwidth]{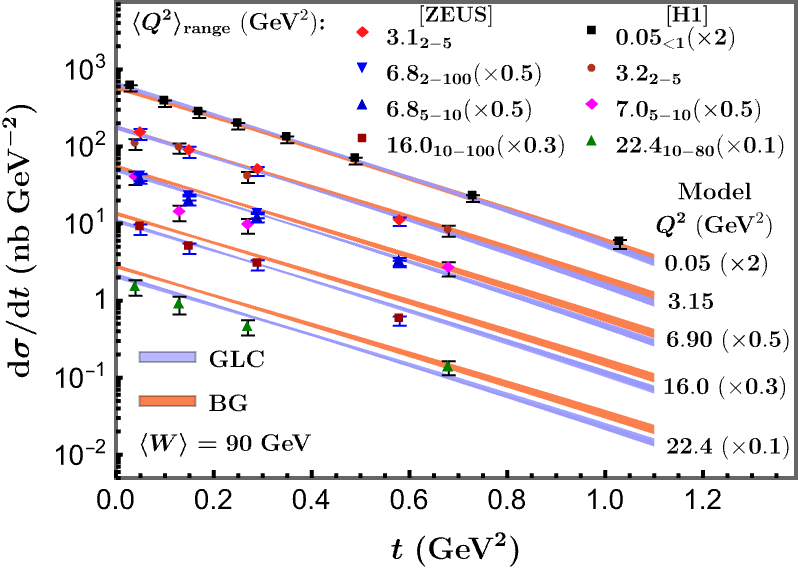}}
\subfigure{\includegraphics[width=0.475\textwidth]{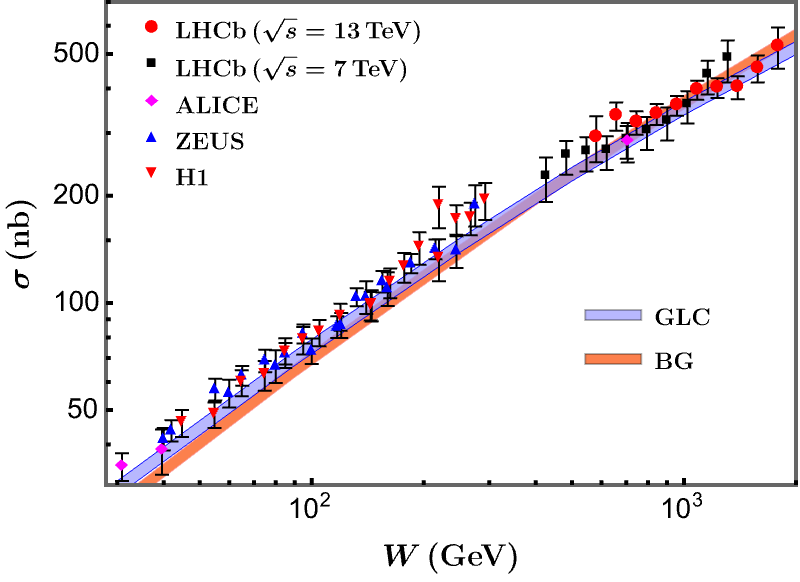}}
\caption{Left: GTMD model fit of the $t$ dependence of exclusive coherent diffractive $J/\psi$ production data from H1 \cite{H1:2005dtp} and ZEUS \cite{ZEUS:2004yeh}, for two vector meson wave function models: GLC and BG. The systematic and statistical uncertainties are added in quadrature. The bands correspond to the ranges of values of $\bar{\chi}$ and $R_p$ specified in the text. Right: Fit of the model to the $W$ dependence of the total $J/\psi$ photoproduction cross section at HERA and LHC \cite{H1:2000kis, ZEUS:2002wfj,ALICE:2012yye,LHCb:2018rcm,LHCb:2014acg}. The data are only shown for $x\lesssim 0.01$ which corresponds to $W>30$ GeV ($\approx 10 M_{J/\psi}$).} 
\label{JPsi_BG-LC}
\end{figure}

\begin{figure}[htb]
\centering
\subfigure{\includegraphics[width=0.45\textwidth]{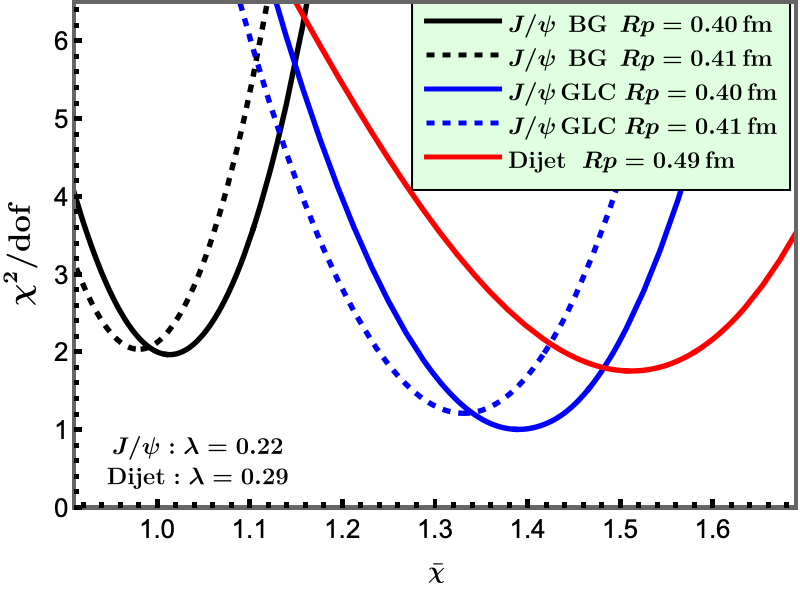}}
\subfigure{\includegraphics[width=0.45\textwidth]{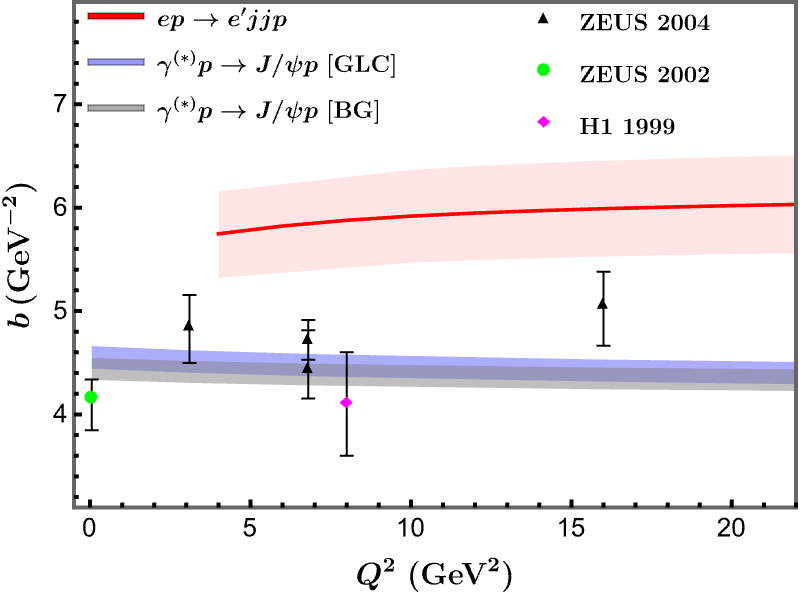}}
\caption{Left: $\chi^2/\text{dof}$ vs $\bar{\chi}$ for dijet production (combined $Q^2$, $t$, $K_\perp$, and $y$ dependence data \cite{H1:2011kou} with dof=15) and $J/\psi$ production (combined photo- and electroproduction data \cite{H1:2005dtp,ZEUS:2004yeh} with dof=35) for two different possible values of $R_p$. Right: $t$ slope ($b$) of the model as a function of $Q^2$ for the dijet and $J/\psi$ cases. The $J/\psi$ data are taken from \cite{H1:1999ujo,ZEUS:2004yeh,ZEUS:2002wfj}. Note that the diffractive dijet production case is $y$ integrated, while the $J/\psi$ case is evaluated at $W=90 \, \text{GeV}$ and for fixed $y$.}
\label{Jpsi_Dijet}
\end{figure}

In Fig.\ \ref{JPsi_BG-LC} we show that our MV-like model with free fit parameters $\bar{\chi}$, $R_p$ and $\lambda$ and without the mentioned corrections can achieve a good description of the $t$ distribution data of HERA \cite{H1:2005dtp,ZEUS:2004yeh} and of the data on the total cross section as a function of $W$ from HERA (H1 \cite{H1:2000kis} and ZEUS \cite{ZEUS:2002wfj}) 
and LHC (ALICE \cite{ALICE:2012yye} and LHCb \cite{LHCb:2018rcm,LHCb:2014acg}). The $t$ dependence of the model is controlled by the proton profile $R_p$, the $W$ slope by $\lambda$, while the amplitude of the cross section is $\bar{\chi}$ dependent where large $\bar{\chi}$ means a larger cross section. Here we give preference to the description of the photoproduction data, i.e.\ the parameters are obtained from a simultaneous description of the $t$ and $W$ dependence of the photoproduction data, where there is only a very narrow range of $R_p$ and $\bar{\chi}$ values that describe those data simultaneously. If one would include electroproduction data, the parameters would change considerably (cf.\ Fig.~\ref{Jpsi_Dijet} (left)) such that the photoproduction data would be described less well. Here we note that the H1 and ZEUS data sometimes differ quite a bit from each other, when comparing data sets at the same or very similar values of $Q^2$. Therefore, we prefer to let the $Q^2$ dependence be determined by the model after the parameters are fixed at $Q^2=0.05\, \text{GeV}^2$. It then turns out that the $t$ dependence is well described using GLC for all values of $Q^2$ considered, while BG overestimates the data for large $Q^2$.

We show bands of values of $\bar{\chi}$ and $R_p$ which describe the photoproduction data qualitatively equally well within the errors of the data points. However, they should not be interpreted as 1$\sigma$ error bands, as they are not obtained from a fit to all data points through a minimization w.r.t.\ all parameters simultaneously. Rather, we determine $R_p$ from the $t$ dependence, $\lambda$ from the $W$ dependence, and subsequently obtain $\bar{\chi}$. Given the uncertainties in the data and the tension with the dijet data, a more sophisticated determination of the parameters and the error bands does not seem to be called for at this point. To be specific about the bands, for GLC we use $\bar{\chi}=1.45-1.50$ for $R_p=0.40$ fm and $\bar{\chi}=1.40-1.45$ for $R_p=0.41$ fm, while for BG we use $\bar{\chi}=1.10-1.15$ for $R_p=0.40$ fm and $\bar{\chi}=1.05-1.10$ for $R_p=0.41$ fm.
We choose two different possible $R_p$ because of the fact that the photoproduction data prefers a steeper slope ($R_p=0.41$ fm) than the electroproduction data ($R_p=0.40$ fm). The parameter $\lambda$ is determined by the slope of the $W$ dependence data of the $J/\psi$ photoproduction total cross section from HERA and LHC in Fig.~\ref{JPsi_BG-LC} (right), to which it is very sensitive. It turns out that unlike the dijet case this data prefers $\lambda=0.22 <\lambda_{\text{GBW}}$. We note that trying to obtain a better fit of the $W$ dependence of the total cross section will lead to a less good description of the photoproduction differential cross section $d\sigma/dt$. As the model is probably less appropriate for the integral over all $t$, we have given preference to the latter. We refer to \cite{Mantysaari:2022sux} for a combined description of H1 data of total photoproduction cross section and the differential cross section for both coherent and incoherent diffraction, taking into account proton shape fluctuations.

In Fig.~\ref{Jpsi_Dijet} (left) the $\chi^2$ per degree of freedom (dof) as a function of $\bar{\chi}$ is shown for the case that a fit is made to both the photo- and electroproduction data. It can be seen that also in this case GLC gives a slightly better minimal $\chi^2/\text{dof}$ and prefers a $\bar{\chi}$ close to the one of the dijet case. Therefore, the GLC model seems to be preferred. However, when it comes to the $t$ and $W$ slope, all vector meson wave function models require $\bar{\chi}$, $R_p$ and $\lambda$ values that are smaller than those obtained from the dijet data. This is clearly visible in Fig.~\ref{Jpsi_Dijet} (right) where we show the value of $b$ of $d^2 \sigma / dt \, dQ^2 \propto e^{-bt}$ for $ep$ collisions, where $b$ is solely determined by $R_p$ in Eq.~(\ref{proton_profile}). The preferred proton profile for the $J/\psi$ case has $R_p=0.40-0.41$ fm, while for the dijet case it is $R_p=0.49$ fm. Here it should be recalled that the $J/\psi$ data is for fixed $y$ ($\gamma^{(*)}p$) and the dijet data is $y$ integrated ($ep$) (cf.\ the bottom right plot of Fig.\ \ref{DifDijet} for the $y$ dependence of the dijet data). The $d^2\sigma/dt dQ^2$ slope data points in Fig.~\ref{Jpsi_Dijet} (right) are only available for $J/\psi$ production \cite{H1:1999ujo,ZEUS:2004yeh,ZEUS:2002wfj}, while the $Q^2$ dependence of the dijet slope is extracted from the fit (see Fig.~\ref{DifDijet}). The dijet slope data is given only for $Q^2$ and $y$ integrated which is $b=5.89\pm 0.50 \, \text{GeV}^{-2}$. The bands of the $t$ slope ($b$) for the $J/\psi$ case shown in Fig.~\ref{Jpsi_Dijet} (right) reflect the aforementioned $R_p$ values, while for the dijet case the bands correspond to the $1\sigma$ error in $R_p$, i.e.\ $R_p=0.49\pm 0.02$ fm, for which a larger $R_p$ gives a larger $b$.
Our model shows that the dijet slope is slowly increasing in $Q^2$ while for $J/\psi$ it is steadily decreasing. This tension cannot be resolved with the present set of just three free parameters.
It is also clear that it cannot be attributed to the vector meson wave function, but stems from the proton profile. Of course, it may be (in part) due to the aforementioned caveats about the dijet data, but without additional future data, that can likely not be clarified.  

\begin{figure}[htb]
\centering
\subfigure{\includegraphics[width=0.45\textwidth]{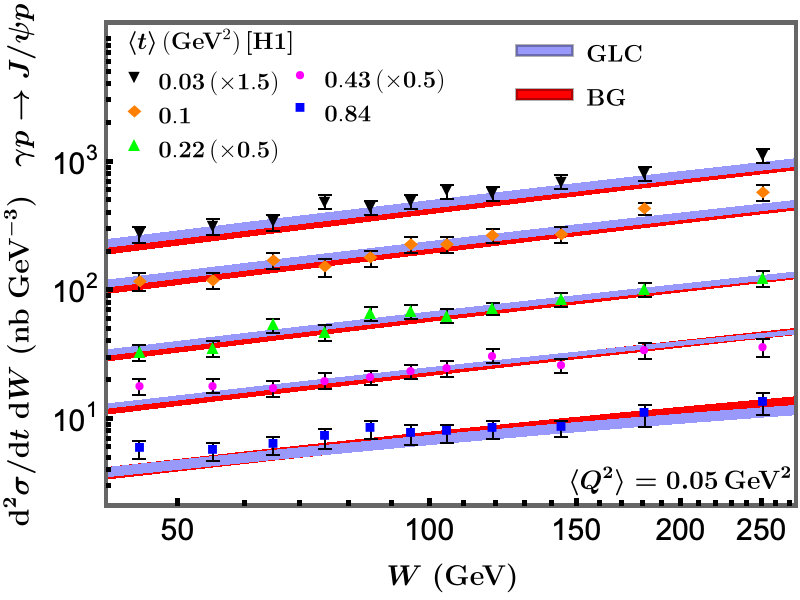}} 
\subfigure{\includegraphics[width=0.45\textwidth]{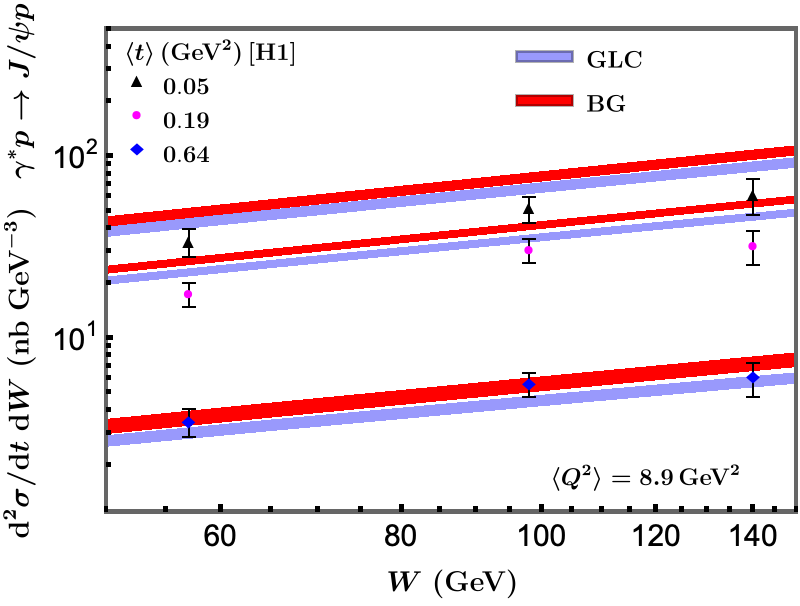}} 
\caption{$W$ dependence of the model fits compared to the H1 data for photoproduction (left panel) and for $\langle Q^2\rangle = 8.9$ GeV$^2$ (right panel) for the two different wave function models and the same model parameters as in Fig.~\ref{JPsi_BG-LC}. }
\label{W_dep}
\end{figure}

In Fig.\ \ref{W_dep} we show the $W$ distribution resulting from the model fits to the $t$-dependence. Both models can describe well the small $Q^2$ data (photoproduction). For electroproduction GLC overestimates the data by at most $2\sigma$, while BG shows a larger deviation from the data, as shown in Fig.~\ref{W_dep} (right). 
 					
\begin{figure}[htb]
\centering
\subfigure{\includegraphics[width=0.45\textwidth]{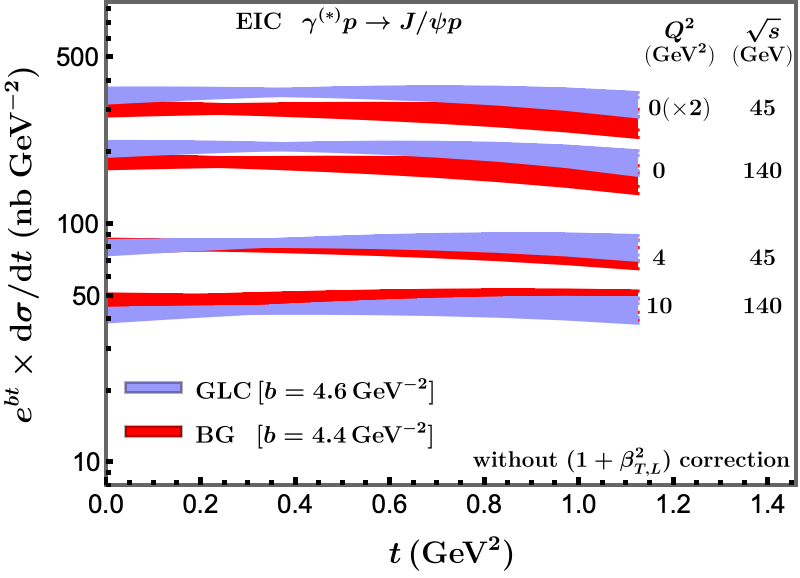}} 
\subfigure{\includegraphics[width=0.45\textwidth]{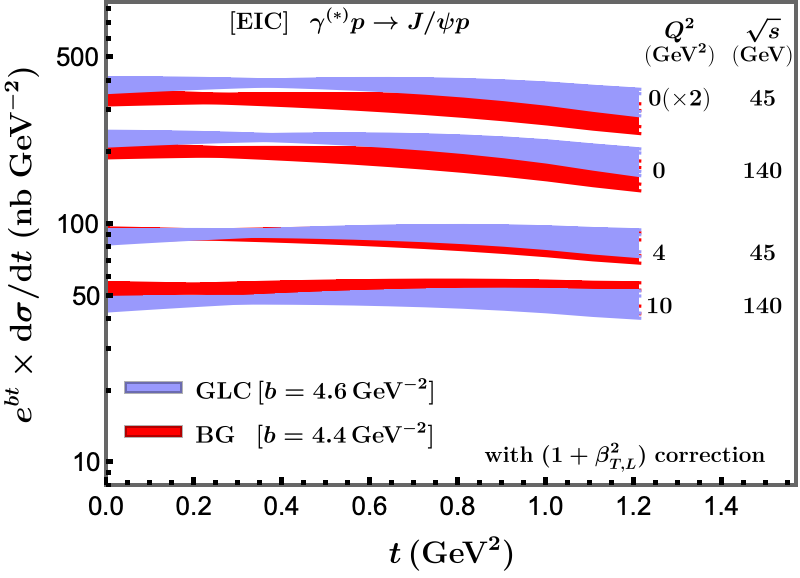}}
\caption{Predictions of diffractive $\gamma^{(*)}p\rightarrow J/\psi\, p$ at the EIC for two different wave function models, without $\beta$ correction (left) and with $\beta$ correction (right). We choose $W=40$ GeV for $\sqrt{s}=45$ GeV which will probe $y\approx 0.79$ while $W=50$ GeV for $\sqrt{s}=140$ GeV will probe $y\approx 0.13$, where we use $y=\left(Q^2+W^2 \right)/s$.}
\label{EIC_pred}
\end{figure}

\begin{figure}[htb]
\centering
\subfigure{\includegraphics[width=0.45\textwidth]{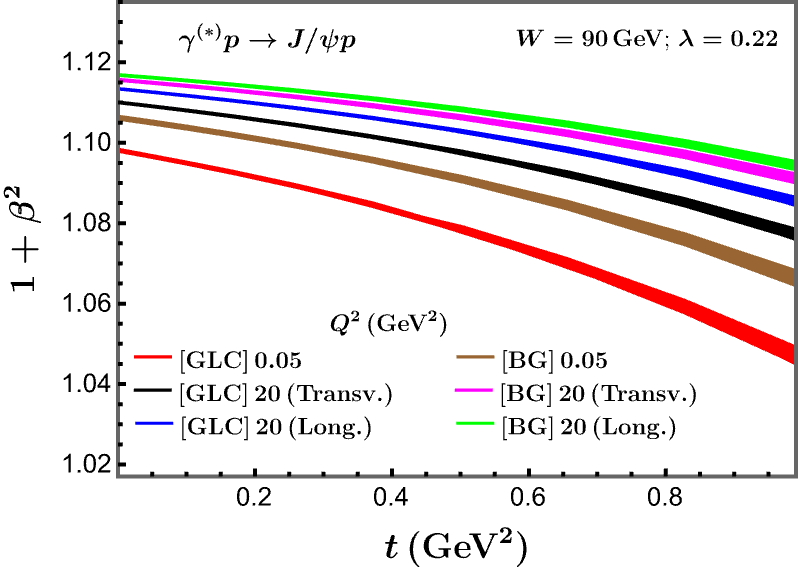}} 
\subfigure{\includegraphics[width=0.45\textwidth]{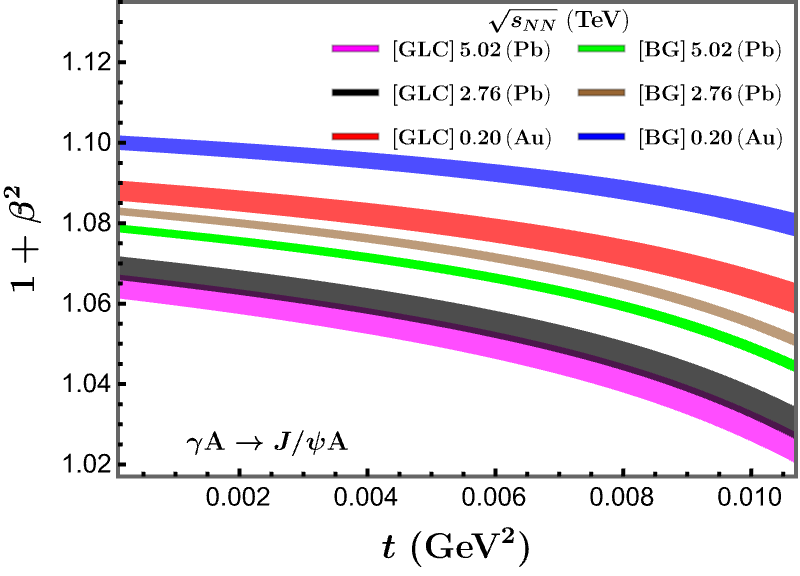}} 
\caption{The phenomenological correction $1+\beta^2$ of the model for diffractive $J/\psi$ production in $\gamma^{(*)}p$ (left) and $\gamma \text{A}$ in UPCs (right). We show only small $t$ for $\gamma \text{A}$ because the correction diverges at the diffractive dip.}
\label{beta_corrct}
\end{figure}

With the obtained model fits we provide predictions for the same process at EIC, which will cover a different kinematic region, but not so different that evolution will play a big role. The left panel of Fig.~\ref{EIC_pred} shows the predictions without the phenomenological correction $1+\beta^2$ and the right panel with. In Fig.~\ref{beta_corrct} we show the correction by itself.   
				
 In the left panel of Fig.\ \ref{beta_corrct} the correction obtained with the model is plotted for $\gamma^{(*)}p$ collisions as a function of $t$ and found to be in the 10-15\% range, a bit larger in the BG model than in the GLC model, and slowly increasing in $Q^2$. This is in agreement with other phenomenological studies \cite{Martin:1999wb,Toll:2012mb,Mantysaari:2016jaz}. For longitudinal photon polarization a similar size correction is obtained. The right panel shows similar size corrections for UPCs to which we will turn next.
		
\section{Coherent Diffractive $J/\psi$ Production in UPCs at midrapidity}
\label{sec:JPsi_pred}
		
Ultra-peripheral Heavy Ion Collisions at LHC and RHIC can be used to study photon-nucleus collisions. In order to make sure that one is dealing with exclusive coherent diffractive production of a $J/\psi$ there need to be rapidity gaps between the $J/\psi$ and both nuclei. We consider the $J/\psi$ to be produced at mid-rapidity. The central rapidity range of the LHC is $|Y|<0.8$ and at RHIC $|Y|<1$. Here we simply take $Y=0$. In that case the colliding photon and gluon will both have an energy of $M_V/2$, which means that the gluon has a momentum fraction of $x_g = M_V/\sqrt{s_{NN}}$. For the LHC Pb-Pb UPC data \cite{ALICE:2021tyx} at $\sqrt{s_{NN}}=5.02$ TeV (Run 2) this corresponds to $x_g=6\cdot 10^{-4}$ and at $\sqrt{s_{NN}}=2.76$ TeV (Run 1) to $x_g=0.001$. For the RHIC Au-Au UPC data at $\sqrt{s_{NN}}=200$ GeV this corresponds to $x_g=0.015$, which is at the edge of the range of applicability of the MV-like model that we are using. We will use the ALICE data to fit $\eta$ and then obtain predictions for RHIC, keeping in mind this caveat.
		
In order to compare the UPCs to the photoproduction case at EIC, it is useful to know what is $W_{\gamma \text{N}}$ in the UPCs. For A-A UPCs at mid-rapidity ($Y=0$), the photon-nucleon center of mass energy squared is determined by $W_{\gamma \text{N}}^2=M_V \sqrt{s_{NN}}$ \cite{Mantysaari:2017dwh}. For LHC Pb-Pb UPC data at $\sqrt{s_{NN}}=5.02$ TeV this corresponds to $W_{\gamma \text{N}}=125$ GeV and at $\sqrt{s_{NN}}=2.76$ TeV to $W_{\gamma \text{N}}=93$ GeV, while for RHIC Au-Au UPC data at $\sqrt{s_{NN}}=200$ GeV this corresponds to $W_{\gamma \text{N}}=25$ GeV.
	
\begin{figure}[htb]
\centering
\subfigure{\includegraphics[width=0.45\textwidth]{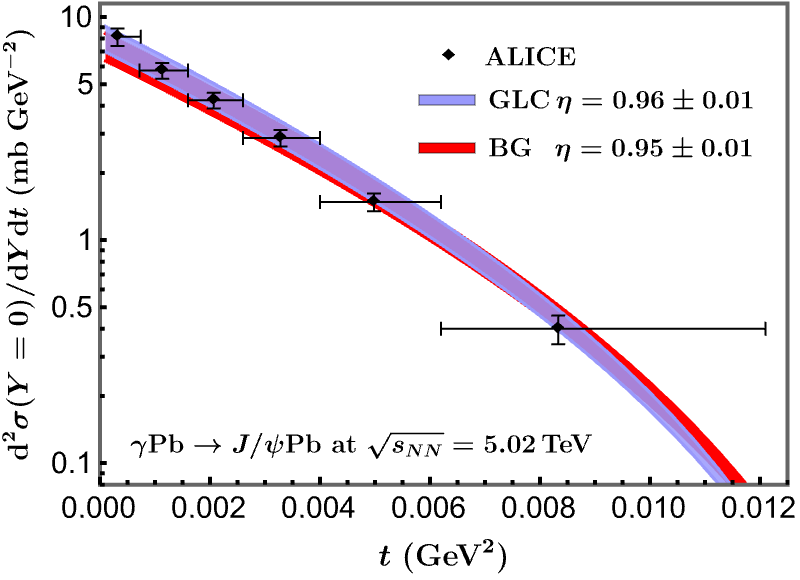}} 
\subfigure{\includegraphics[width=0.45\textwidth]{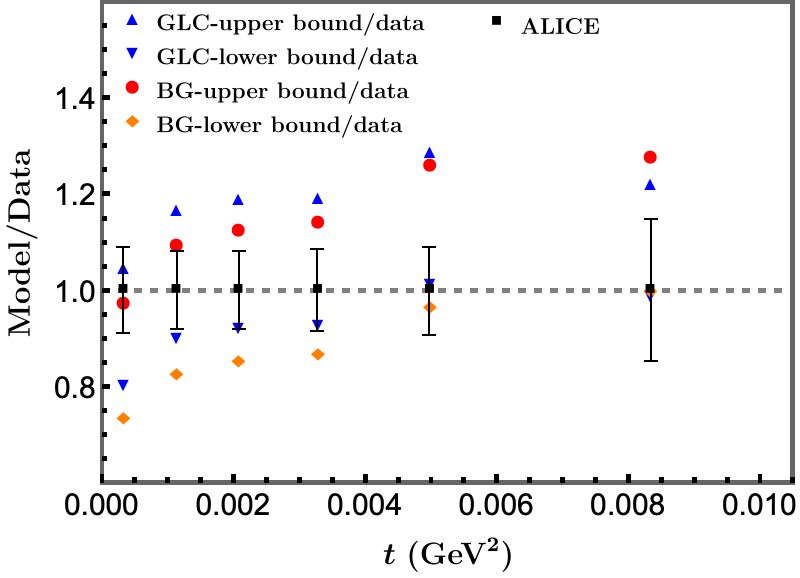}} 
\caption{(left) Fit of the model to ALICE (Run 2) data \cite{ALICE:2021tyx} of coherent diffractive $J/\psi$ production at midrapidity in ultra-peripheral Pb-Pb collisions at $\sqrt{s_{NN}}=5.02$ TeV. We use the same parameter values as in Fig.~\ref{JPsi_BG-LC}, but fit the additional parameter $\eta$ that determines the saturation scale for nuclei.  (right) Ratio of the model fit to the ALICE data for lower and upper bounds of each wave function.
}
\label{ALICE_fit}
\end{figure}

In Fig.~\ref{ALICE_fit} we provide a fit of our model to the $t$ dependence of the differential cross section $d^2\sigma/dYdt$ \cite{ALICE:2021tyx} for coherent diffractive $J/\psi$ production at $Y=0$ in ultra-peripheral Pb-Pb collisions at $\sqrt{s_{NN}}=5.02$ TeV. 
According to \cite{ALICE:2021tyx}, various models can describe the ALICE data well. One such model is the leading-twist approximation (LTA) of nuclear shadowing \cite{Frankfurt:2011cs,Guzey:2016qwo}, which combines the Glauber-Gribov formalism with the phenomenology of photon diffraction from HERA. The lower bound of the GLC fit in our model is close to the LTA (low shadowing) model result with a slightly steeper slope, as shown in Fig.~\ref{ALICE_fit} (right). Another model, the b-BK \cite{Cepila:2020xol,Bendova:2019psy,Bendova:2020hbb}, was proposed based on the solution of the Balitsky-Kovchegov equation \cite{Balitsky:1995ub,Kovchegov:1999yj} with an impact parameter dependence. Another study that incorporates nucleon shape fluctuations \cite{Mantysaari:2022sux} also provides a good fit to the data, including the coherent $W$ dependence of the photoproduction total cross section and the $t$ distribution of coherent and incoherent $J/\psi$ photoproduction data from HERA. While the former two models utilize the BG wave function, our model provides a better description of the $t$ and $W$ dependence data using the GLC wave function. Another wave function model based on the Buchm\"{u}ller-Tye potential \cite{Buchmuller:1980su} that uses $\vec{r}$-$\vec{b}$ correlation \cite{Kopeliovich:2021dgx} with two different parameterizations: GBW \cite{Golec-Biernat:1998zce,Golec-Biernat:1999qor} and BGBK \cite{Bartels:2002cj}, also gives good agreement with the data \cite{Kopeliovich:2022jwe}. Differences in the magnitude and slope of the cross section between other models and ours could also be due to the use of different nuclear radius parameters.

Extrapolating the fit gives a prediction of the first diffractive minimum (or dip) to be at $t\simeq 0.016 \, \text{GeV}^2$. We find that the dip position is determined by the target profile $R_A$, such that it will move towards a smaller $t$ value for larger $R_A$. The fit turns out to be very sensitive not only to the value of $R_A$, but also to the power of $A$. With the fit of the model to the ALICE data, we find that $\eta=0.96 \pm 0.01$ for GLC and $\eta=0.95 \pm 0.01$ for BG give the best fit. Therefore, our model fit indicates that the saturation scale behaves like $Q_s^2 \propto A^{0.27-0.30}$, not $A^{1/3}$. The latter in fact does not provide a good fit. Following Eq.~(\ref{Saturation2}), the saturation scale $Q^2_{0s,\text{A}}$ for the heavy nucleus $A$ depends on $\eta$, $R_A$, and $R_p$, and is in all cases found to be between 1.5 and 1.9 $\text{GeV}^2$. The choice of proton and nuclear profile, particularly the Gaussian profile used in our study, can affect the fitted $\eta$ value and may differ with different profiles. We did not attempt to find profiles that would lead to $\eta =1$ and do not exclude that that is possible.

\begin{figure}[htb]
\centering
\subfigure{\includegraphics[width=0.325\textwidth]{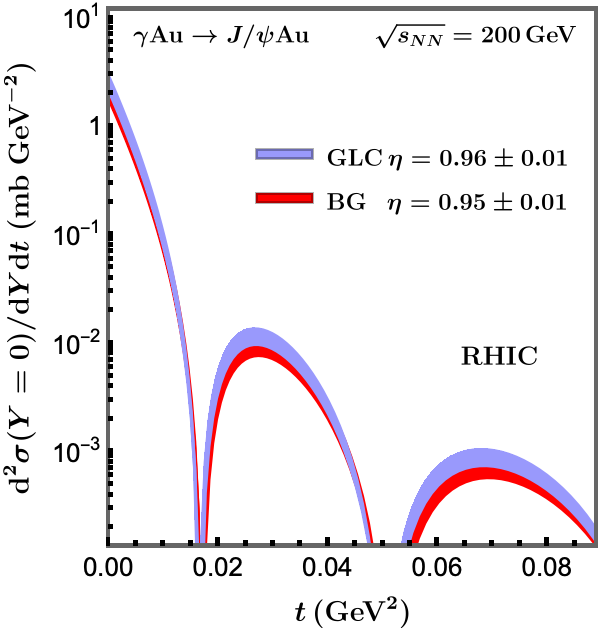}} 
\subfigure{\includegraphics[width=0.325\textwidth]{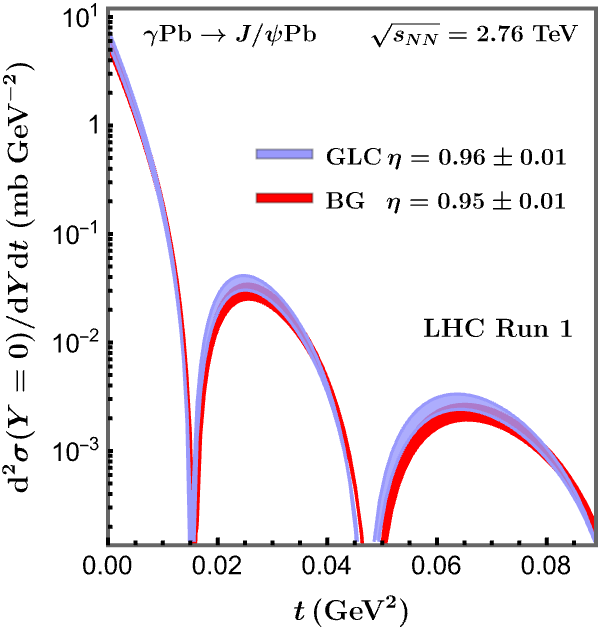}} 
\subfigure{\includegraphics[width=0.325\textwidth]{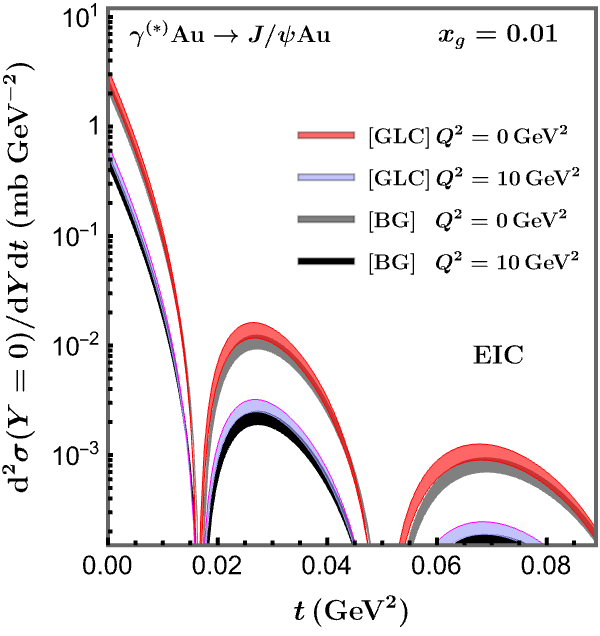}} 
\caption{Model predictions of coherent diffractive $J/\psi$ production at midrapidity in ultra-peripheral Au-Au collisions with $\sqrt{s_{NN}}=200$ GeV at RHIC (left) and Pb-Pb collisions with $\sqrt{s_{NN}}=2.76$ TeV at LHC (middle). We also give predictions for $e$-Au collisions at the EIC (right) at fixed $x_g=0.01$ at $Q^2=0\, \text{GeV}^2$ and $Q^2=10\, \text{GeV}^2$.}
\label{Predictions2}
\end{figure}

With the same parameterization, in Fig.~\ref{Predictions2} we provide predictions for RHIC and LHC (Run 1) at midrapidity. In general, both wave function models give very similar results but GLC shows a somewhat larger band than BG as expected from the previous analysis on $\gamma p$ (see the $\chi^2/\text{dof}$ in Fig.~\ref{Jpsi_Dijet}). The left panel in Fig.~\ref{Predictions2} is for Au-Au UPCs at $\sqrt{s_{NN}}=200$ GeV. Fig.~\ref{Predictions2} shows that the first diffractive dip for Au-Au at RHIC is predicted to be around $t\simeq 0.017\pm 0.001 \, \text{GeV}^2$ which is close to its location in RHIC preliminary data \cite{Adam2020} and other studies \cite{Cepila:2017nef,Mantysaari:2017dwh}, while for Pb-Pb at LHC Run 1 it is predicted to be around $t\simeq 0.015\pm 0.001 \, \text{GeV}^2$. The middle panel in Fig.~\ref{Predictions2} is for Pb-Pb UPCs at LHC Run 1 
with $\sqrt{s_{NN}}=2.76$ TeV. In the rightmost panel of Fig.~\ref{Predictions2}, we provide predictions of $e$-Au collisions at the EIC for photoproduction ($Q^2=0\, \text{GeV}^2$) and electroproduction ($Q^2=10\, \text{GeV}^2$) for fixed $x_g=0.01$ which corresponds to 
$W_{\gamma \text{N}}=31\, \text{GeV}$ 
and 
$W_{\gamma \text{N}}=44\, \text{GeV}$, respectively.
The bands for each wave function model reflect the uncertainties on $R_p$ (which translates to $\eta$) and $\bar{\chi}$. As shown in Fig.~\ref{beta_corrct}, the $\beta$ corrections are in the order of 6-10\% for UPCs, which is small compared to the uncertainties, hence not included here.
 
\section{Discussion and Conclusions}
\label{sec:discussion}
	
We have shown that incorporating $x$ dependence in our previous gluon GTMD model, along the lines of the GBW parameterization of the saturation scale, will improve the description of HERA-H1 diffractive dijet production data. However, describing diffractive $J/\psi$ production in the same way leads to tension for the slope of the $t$ dependence, which is quite distinct for the two processes, where dijet needs a steeper slope than $J/\psi$ production. In the model this slope is solely determined by the (Gaussian) proton profile and there appears to be no clear way to resolve the tension. A few other differences between the optimal parameter choices for dijet and $J/\psi$ production are found, but these can be reduced by adjusting the $J/\psi$ wave function or by modifying the $x$ dependence w.r.t.\ the GBW parameterization. For instance, dijet production can be described well by an $x$ dependence of the saturation scale $Q_s \propto x^{-\lambda}$ with $\lambda=\lambda_{\text{GBW}}=0.29$, while the $W$ dependence of photoproduction of $J/\psi$'s prefers a smaller $\lambda \approx 0.22$. A non-constant $\lambda$ may be needed, but we did not explore that option, anticipating that future more precise data may shed new light on the differences between dijet and $J/\psi$ production. We do expect that a common gluon GTMD model description of dijet and $J/\psi$ production may be possible once the slope issue is clarified by new data.

With the best fit of our model to combined H1 and ZEUS data on $J/\psi$ production, we have provided predictions for the diffractive $J/\psi$ production in $e$-$p$ collisions at the future EIC. We further fit our model to UPC data from ALICE (Run 2) to determine the $A$ dependence of the saturation scale. The fit turns out to be very sensitive to the power of $A$ and the nuclear profile $R_A$, which suggests that also the profile function shape will matter considerably. We find an $A$ dependence that is slightly less than the generally expected $A^{1/3}$, to be specific, $Q_s^2 \propto A^{0.27-0.30}$. With the obtained fit we provide predictions for UPCs at LHC (Run 1) and RHIC, and for $e$-Au collisions at EIC. We have also investigated a phenomenological correction commonly used for $J/\psi$ production which comes from the odderon contribution, which is in the 10-15\% range and thus unimportant given the present large uncertainties in the available data and hence in the model. The larger correction from accounting for non-zero skewness that is often considered in GPD approaches in the DGLAP regime does not seem appropriate for our GTMD model and was thus not included. Moreover, a similar correction has not been applied in dijet production, which would affect the comparison. We have also pointed out that dijet and $J/\psi$ production go beyond probing GPDs, rather they probe weighted integrals of GTMDs with weights that depend on external kinematical variables of the process that can be varied and exploited.

The predictions from the presented $x$-dependent gluon GTMD model for $J/\psi$ production at EIC, LHC, and RHIC, will hopefully facilitate resolution of the $t$ distribution tension with dijet production, clarify the dependence on the skewness probed in the process, and determine the $x$ and $A$ dependence of the saturation scale of heavy nuclei.

\acknowledgments
We thank Gerco Onderwater for useful discussions on the fits and Cristina S\'anchez Gras on the LHCb data. We thank the Center for Information Technology of the University of Groningen for their support and for providing access to the Peregrine high performance computing cluster. The work of C.S. was supported by the Indonesia Endowment Fund for Education (LPDP).

\end{document}